\documentclass[10pt,letterpaper]{article}
\usepackage{opex3}
\usepackage{amsmath}
\usepackage{cite}
\begin{document}

%%%%%%%%%%%%%%%%%% title page information %%%%%%%%%%%%%%%%%%
\title{Space-time resolved simulation of femtosecond nonlinear light-matter interactions using a holistic quantum atomic model
: Application to near-threshold harmonics}

\author{M.~Kolesik,$^{1,2,*}$ E.~M.~Wright,$^{1}$ J.~Andreasen,$^{1}$ J.~M.~Brown,$^{1}$
        D.~R.~Carlson,$^{1}$ and R.~J.~Jones$^{1}$}

\address{
$^1$College of Optical Sciences, University of Arizona, Tucson, AZ 85721, U.S.A.\\
$^2$Department of Physics, Constantine the Philosopher University, Nitra, Slovakia
}

\email{$^*$kolesik@acms.arizona.edu} %% email address is required



\begin{abstract}
We introduce a new computational approach for femtosecond pulse propagation in the transparency region of gases that
permits full resolution in three space dimensions plus time while fully incorporating quantum coherent effects such
as high-harmonic generation and strong-field ionization in a holistic fashion.  This is achieved by utilizing a
one-dimensional model atom with a delta-function potential which allows for a closed-form solution for the nonlinear
optical response due to ground-state to continuum transitions. It side-steps evaluation of the wave function, and
offers more than one hundred-fold reduction in computation time in comparison to direct solution of the atomic Schr\" odinger
equation. To illustrate the capability of our new computational approach, we apply it to the example of near-threshold
harmonic generation in Xenon, and we also present a qualitative comparison between our model and results from an
in-house experiment on extreme ultraviolet generation in a femtosecond enhancement cavity.
\end{abstract}


%%%%%%%%%%%%%%%%%%%%%%% References %%%%%%%%%%%%%%%%%%%%%%%%%

%\bibliographystyle{osajnl}
%\bibliography{hhgwithoda}

%%%%%%%%%%%%%%%%%%%%%%%%%%  body  %%%%%%%%%%%%%%%%%%%%%%%%%%
\section{Introduction}

The goal of this paper is to present a new computational approach for nonlinear light-matter interactions in atomic and molecular gases
for incident femtosecond pulses whose spectra lie in the transparency region, the approach opening the door to fully resolved
propagation in three spatial dimensions plus time, or (3+1) dimensions.  For sufficiently low input intensities when ionization of
the constituent atoms is negligible and the atoms remain dominantly in their ground state, suitable space-time resolved computational
approaches already exist and incorporate the linear dispersive properties of the medium, both refractive-index and absorption,
along with a model for the bound state electronic nonlinearity, typically a Kerr-type nonlinearity possibly with saturation,
and also a time-delayed nonlinear response to model Raman effects for molecules~\cite{Couairon2007}.
For higher intensities when atomic ionization becomes relevant a number of effects turn on that involve ground state to
continuum state transitions. 
These transitions can 
greatly modify field propagation via nonlinear generation of freed electrons and associated
nonlinear losses and refraction, and high harmonic generation (HHG). Typically the nonlinear optical response arising from ground
state to continuum transitions are dealt with in a piecemeal fashion. For example, the polarization source term for high harmonics
is often calculated using the strong-field approximation~\cite{LewensteinPRA94} whereas the rate of ionization is treated in a
separate calculation using Keldysh theory or one of the subsequent variants such as ADK theory (see~\cite{BergeRPP07} for a
useful compendium of different ionization-rate formulas).
Such an approach has met with considerable success to date for simulations
of HHG (ref.~\cite{Gaarde2008} has a readable yet detailed description of the method)
and also light filament propagation in gases~\cite{Couairon2007},
but it would clearly be advantageous for the advancement of these fields if the HHG and all ionization related effects could be modeled
in a holistic manner so that as parameters are varied the relative roles of high-harmonic generation and ionization related effects
are microscopically consistent without having to tune the parameters of separate models. An obvious candidate for such a quantum
model is to solve the three-dimensional (3D) atomic Schr\" odinger 
equation~\cite{nurhuda_generalization_2008,Volkova2011} 
to obtain the quantum averaged dipole moment at each point
in space, and use this as a source in Maxwell's equation for the field propagation.  However, due to the restrictive computing
requirements needed to solve the 3D Schr\" odinger equation
% at a large number of space points for fully space-time resolved field propagation
it seems clear that this will not be a viable option for the foreseeable future.

The  quantum model we employ in the current work is one-dimensional, and it is perhaps the simplest system which has both the
energetic continuum and a bound state --- the minimal set of ingredients crucial to describe light-matter interactions at
femto-second time scales and ionization-level intensities.
However, the model has the virtue that it can be implemented to efficiently simulate space-time resolved field propagation
in (3+1) dimensions. To the best of our knowledge Christov~\cite{Christov2000} was the first to perform simulations of femtosecond
pulse propagation in one space dimension plus time, or (1+1) dimensions, that incorporated the medium response via direct numerical
solution of a time-dependent 1D atomic model.
%, and he investigated HHG and attosecond pulse generation.
In a series of recent papers Bandrauk and co-workers~\cite{lorin_wasp_2011} have presented simulations of 1D quantum models
coupled to field propagation
in both (1+1) and (3+1) dimensions, and addressed a range of topics of current interest including HHG and filamentation in gases.
One-dimensional atomic models have played an important role as model systems for elucidating the physics underlying nonlinear
light-matter interactions. For example, Su and Eberly~\cite{su_model_1991} introduced the quasi-Coulomb potential in 1D as a model for
multi-photon physics.
Here we use the $\delta$-potential atomic model. This model has one free parameter, the ionization potential,
and its spectrum consists of one bound state, the ground state, plus the continuum, so it is an ideal
model system in which to explore the role of ground state to continuum transitions. Indeed, this system has been
studied and applied in various contexts for many years. The attraction of the model as a toy system to investigate
various aspects of dynamics in quantum systems stems from the fact that many quantities can be obtained
exactly~\cite{arrighini_ionization_1982,elberfeld_tunneling_1988}.
Its generalizations to higher dimensions~\cite{cavalcanti_decay_2003},
for more complex potentials~\cite{uncu_solutions_2005,alvarez_perturbation_2004},
and other dynamic equations~\cite{villalba_particle_2009} also exist.
Besides its utility in the AMO and strong field
physics~\cite{Geltman1978,dunne_simple_2004,su_stabilization_1996,dziubak_stabilization_2010},
the 1D $\delta$-potential
has also been applied in semiconductors~\cite{elberfeld_tunneling_1988,uncu_solutions_2005}, and
used as a test system for numerical and approximate
methods~\cite{sanpera_ionization_1993,geltman_short-pulse_1994,zhao_boundary-free_2002}.
We have applied the $\delta$-potential model to explore the relative roles of the higher-order Kerr effect and plasma induced
defocusing in filament propagation~\cite{teleki2010,Brown2012}.
% and also evaluated the high-harmonic spectrum.
Furthermore, it has recently been shown by two
of the present authors (MK and JMB) that the $\delta$-potential model yields a closed form solution for the quantum averaged
nonlinear current for this model atom in an arbitrary time-dependent light field~\cite{Brown2011}.
(The code computing the nonlinear current  is available as open-source upon request to the
corresponding author (MK).)  This exact solution side steps the need for direct evaluation of the time-dependent atomic
wave function and therefore represents an enormous savings in computational effort and time.  For example, one does not need
to worry about griding issues for the solution of the 1D wave equation and the spurious effects of numerical boundary conditions.
Our new approach is computationally efficient since it employs the closed form for the nonlinear current as a source in the (3+1)
field propagation equation, and we find gains of more than one hundred in computation time
in comparison to the approach using integration of the Schr\" odinger equation.

The remainder of this paper is organized as follows: In Sec. 2 we describe our computational approach and discuss the model equations.
In Sec. 3 we include numerical simulations, and for this initial paper we utilize near-threshold HHG
in Xenon as an illustrative example.  We chose near-threshold HHG, with concomitant lower-orders in the harmonic spectrum,
to showcase the capability of the model to cope with an extreme spectral bandwidth in a completely unified treatment.
Unlike the strong-field approximation, our approach does not rely upon identification of the quantum paths that dominate the HHG spectrum,
and is valid also for very low frequencies.
The first part of Sec. 3 describes simulations of HHG in a Xenon gas jet to demonstrate that our model can deal with different
incident focusing conditions and hence phase-matching regimes, and that it reproduces expected qualitative features.
In this respect we do not claim these results are new, particularly since corresponding simulations of HHG in (3+1) already exist
that employ the strong-field approximation, but rather are intended to highlight the capabilities of the approach.
Sec. 3 also includes what we believe are new results to model near-threshold HHG in a femtosecond enhancement cavity (fsEC).
Here a new feature is that the HHG occurs in the presence of the residual plasma that accumulates within the Xenon gas jet due
to the circulating pulse, and we compare the qualitative features of the harmonic spectrum with that of the in-house experiment
that has recently been used to produce record power levels of extreme
ultraviolet (XUV) \cite{Lee11}.

\section{Computational approach and model}

To contrast the model of this paper to the current state of the art in computer simulation of
high-harmonic generation and/or femtosecond filamentation, we briefly recall the main features of the
standard approach. The propagation of the optical and high-harmonic fields are described in
one-way (i.e. directional) propagation equations, into which the various medium responses are
coupled.
For this purpose, we use the Unidirectional Pulse Propagation Equation (UPPE) solver.
In the spectral domain the UPPE for propagation along the z-axis takes the general form \cite{KolesikPRE04}
\begin{equation}
\partial_z E_{k_\perp}(\omega,z) =
i k_z E_{k_\perp}(\omega,z)
+ \frac{i\omega^2}{2 \epsilon_0 c^2 k_z} P_{k_\perp}(\omega,z,\{E\})  -
 \frac{\omega}{2 \epsilon_0 c^2 k_z} J_{k_\perp}(\omega,z,\{E\}) ,
\label{eq:UPPE}
\end{equation}
where the z-component of the optical wavevector is given by
\begin{equation}\label{kz}
k_z(\omega,k_\perp) = \sqrt{\omega^2 (1+\chi(\omega))/{c^2} - k_\perp^2} .
\end{equation}
The UPPE describes the evolution of the spatially resolved spectral amplitudes of the scalar electric field $E_{k_\perp}(\omega,z)$,
where $\omega$ is the frequency, and $k_{\perp}$ denotes transverse wavenumbers, so that the space-time dependent electric
field $E(r,z,t)$ may always be reconstructed using a 3D Fourier (or Hankel) transform over $k_\perp,\omega$.
Note that the assumption of a linearly polarized field is consistent with our adoption of a 1D atomic model.
Nonlinear effects enter the propagation equations through either the
polarization ($P$) or current-density ($J$) terms.  These are both evaluated as functionals of the electric field
history $E(t)$ at each spatial point, and then Fourier-transformed to the spectral representation, e.g. $J_{k_\perp}(\omega,z,\{E\})$,
that appears in the above propagation equation. The linear wavelength-dependent properties of a medium are
represented through the susceptibility$\chi(\omega)$.

In the standard model, there are a number of independent contributions both in the polarization and in the current density.
They include the optical Kerr effect, ionization in strong fields and a freed-electron density evolution equation,
separately modeled losses due to ionization, avalanche ionization, de-focusing effects of freed electrons, and
losses due to freed electrons described in terms of an effective Drude-plasma model~\cite{Couairon2007}.

Our computational approach for simulating femtosecond nonlinear light-matter interactions reflects the fact that
the above mentioned effects are all manifestation of the single electronic system response to the strong optical
field. The goal is to achieve a unified description, and reduce multiple independent model parameters.
In general, our model involves three components for describing an atomic gas:
\begin{itemize}
\item A description of the linear dispersive properties of the gas via a complex-valued frequency dependent susceptibility $\chi(\omega)$.
\item The 1D $\delta$-potential quantum model for modeling the nonlinear current ($J$) due to ground state to continuum transitions.
This will incorporate the effects of ionization, HHG, and ionization induced absorption and refraction in a holistic fashion.
\item A Kerr-like nonlinearity and associated nonlinear polarization $(P)$ to capture the nonlinear optical response due to the ground
state to excited bound state transitions.  This will contribute processes such as four-wave mixing, self-phase modulation, and
self-focusing. (For a molecular gas we may also add a time-delayed nonlinear response to capture Raman effects)
\end{itemize}
For this initial exposition we shall give a description of each of the three components above in the subsections below
as appropriate to our illustrative example of near-threshold HHG in Xenon gas.
The three medium components $\chi,J,P$ are coupled into the field-propagation equation which encompasses all
frequency components from the fundamental to high harmonics.
While we use the Unidirectional Pulse Propagation Equation,  we emphasize that this approach may be implemented
with any pulse-propagation simulator that resolves the carrier wave of the optical field.

\smallskip
{\em Linear dispersive properties}
\smallskip

Inclusion of the linear dispersive properties of the medium over the full span of harmonics is important to properly
incorporate the phase-matching and re-absorption that affect the spatial evolution of the generated harmonics.
In our spectral solver this does not present a problem as long as suitable data is available for the index of refraction and
absorption properties in a sufficiently wide spectral range. The medium is characterized in terms of the linear complex
susceptibility $\chi(\omega)$, and the pulse propagation method utilizes this information in the
propagation constant $k_z(\omega,k_\perp)$ in (\ref{kz}) at
each frequency or wavelength resolved by the numerical simulation.

\begin{figure}[ht]
\centering\includegraphics[clip,width=7cm]{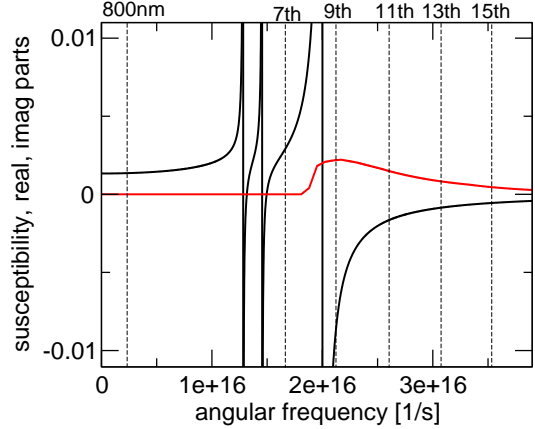}
\caption{
Real (dark lines) and imaginary (red line) parts of the linear susceptibility of Xenon as employed in the
numerical simulations in this paper. The vertical dashed lines
mark the fundamental wavelength of the fundamental field at 800 nm, and
several of its harmonic are also indicated.
}
\label{fig:1}
\end{figure}

Throughout this paper we use Xenon as a representative atomic medium
to showcase the modeling capabilities, and Fig.~\ref{fig:1} shows the
real (solid dark line) and imaginary (red line) parts of the susceptibility as functions of the angular frequency of the
electromagnetic field.
To create the data set displayed in Fig.~\ref{fig:1} we have combined Xenon data downloaded from
{\em http://henke.lbl.gov/optical\_constants/}
with the refractive-index parametrization obtained from Ref.\cite{Xenon} to create a tabulated representation of the linear
susceptibility over a range of frequencies that spans the fundamental at 800~nm wavelength up to its $35^{th}$  harmonic.
%XXX modifications
Note that with the real and imaginary parts of the susceptibility function 
coming from separate sources, and with no absorption data available for 
photon energies below 10~eV, our parametrization does not satisfy 
the Kramers-Kronig relations. To our best knowledge a
more complete data set is currently not available.
Nevertheless,
%XXX modifications
while not a perfect model of the Xenon gas for all frequencies this 
%XXX modifications
data set employed serves a an illustration example in which 
%XXX modifications
%illustrates the point that 
all electromagnetic frequencies
are treated on the same footing by the propagation solver, which operates on the full electric field.  Generally the accuracy
of the linear propagation will be limited by the quality of the available data for the linear susceptibility.

\medskip
{\em Quantum atomic model for the nonlinear response}
\medskip

As alluded to in the introduction our computational approach employs a 1D quantum model with a short-range $\delta$-potential
for the nonlinear current due to ground state to continuum transitions.  More specifically the corresponding time-dependent
Schr\" odinger equation in atomic units takes the form
\begin{equation}
\left[ i \partial_\tau + \frac{1}{2}\partial_s^2 + B\delta(s) - sF(\tau) \right] \psi(s,\tau) = 0 \ \ .
\label{eq:TDSE}
\end{equation}
Here $s$ and $\tau$ are the space and time variables in atomic units, the strength of the attractive potential is governed
by the parameter $B$, with the ionization potential of the ground state equal to $B^2/2$,
and $F(\tau)$ denotes the time-dependent external field in atomic units applied to the atom.

The above Schr\" odinger equation would have to be solved numerically at each spatial point resolved by the
optical pulse simulator. Such an approach, although feasible for this
specific quantum system, is very demanding on computing resources. Fortunately, we do not need the wave function
because it is only the time-dependent dipole moment or current that enters as an input in the UPPE (\ref{eq:UPPE}),
and we have recently shown how an exact, closed form solution can be obtained for both for an arbitrary external
field~\cite{Brown2011}.  More specifically, the solution provides a means of evaluating the expectation value for the
current in atomic units
\begin{equation}
J(\tau) \propto \int_{-\infty}^{\infty}ds\psi^*(s,\tau)\left (i{\partial\over\partial s}\right )\psi(s,\tau) ,
\end{equation}
without having to go through solving Eq. \ref{eq:TDSE} for the wave function $\psi$. Rather, the
current $J$ required for the UPPE is calculated directly, completely by-passing the wave function,
which amounts to a tremendous savings in computing time while retaining the full coherent quantum dynamics
of the atomic system as is pertinent to the nonlinear field propagation. Moreover, it is possible to
exactly eliminate the current component which is linear in the driving field strength~\cite{Brown2011}. 
The quantum model
can therefore be restricted to contribute only to the nonlinear response, which is advantageous when it is
combined with the linear dispersive medium.
For the sake of completeness the formulas for the nonlinear
quantum current are summarized in the Appendix. The reader is referred to \cite{Brown2011}
for details concerning the practical numerical implementation.
For those interested in using this model in their simulations,
the corresponding software is available as an open-source upon request to the corresponding author (MK).

The integration with the pulse propagator solver is in principle the same as for
any other nonlinear medium response. Having calculated the evolution of the optical field
up to a given propagation distance, the history of the electric field at a given point
in space is converted to the external field $F(t)$ in atomic units. This drives
the quantum system, and the induced current is computed as outlined in the Appendix.
The nonlinear current is next converted from the atomic units, and multiplied by the number density atoms
at the point in space. The resulting macroscopic current density is included in the right-hand-side of the UPPE equation.
We remark that since we only incorporate the {\em nonlinear} current from the quantum model into the UPPE,
we do not double-count by erroneously including linear properties from the 1D atomic model.
%XXX modifications
Also note that it is not an option to retain the linear part of the
model's response instead of $\chi(\omega)$ introduced in the previous 
subsection. This is because in a real medium $\chi(\omega)$  originates 
in virtual transitions among a large number of states, and it would be 
difficult to model this from first principles. The linear susceptibility 
arising from our 1D atomic model is far too
simplistic to capture the chromatic properties of a gas to any realistic degree.
%XXX modifications

\begin{figure}[ht]
\centering\includegraphics[clip,width=7cm]{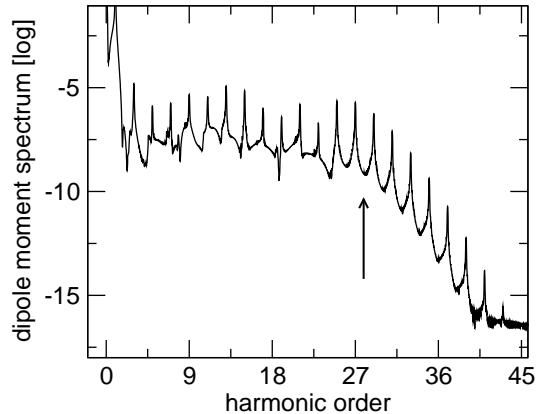}
\label{fig:2}
\caption{
High-harmonics in the spectrum of the dipole moment induced by a
pulse at $\lambda=800$nm. The intensity was $1.5\times 10^{18}$W/m$^2$,
kept constant over duration of ten optical cycles. The arrow marks
the location of cut-off energy calculated for these conditions.
}
\end{figure}

Next we would like to illustrate that our 1D $\delta$-potential atomic model displays features for
HHG that are qualitatively similar to those obtained using the strong field approximation applied to the more
exact 3D Hydrogen-like atomic model. To this end Fig.~\ref{fig:2} shows an example of a harmonic spectrum of
the nonlinear current induced in the 1D model atom for an ionization potential of 12~eV, characteristic of Xenon,
and a ten-cycle pulse of center wavelength $\lambda=800$nm and peak intensity$1.5\times 10^{18}$W/m$^2$. This harmonic
spectrum exhibits the characteristic high-harmonic generation plateau with a high frequency cut-off.
The cut-off predicted by the formula from the standard HHG 
model~\cite{CorkumPRL93,LewensteinPRA94,schafer_above_1993} 
is indicated by the arrow in Fig.~\ref{fig:2} for
our parameters. Furthermore, the harmonic spectrum includes the $9^{th}$ harmonic which occurs near the ionization potential
up to around the $40^{th}$ harmonic, and thus constitutes an example of near-threshold HHG where the generated harmonics straddle
the ionization energy.  What is shown in Fig.~\ref{fig:2} is the spectrum of the nonlinear polarization $P$ that appears in
the Maxwell equations and not the spectrum of the radiation actually generated. Importantly the harmonic spectrum shown is exact,
within the context of our 1D model, over the whole frequency range and in particular at low frequencies.

Finally we point out some pros and cons of our 1D model versus the strong-field approximation.  The strong field approximation
is often utilized in numerical atomic simulations of HHG, and describes the 3D atomic system in terms of a single bound state,
the ground state, and a continuum of free electron continuum states~\cite{LewensteinPRA94}.  In this sense it addresses a similar
spectrum of electronic states as our 1D model, albeit in 3D. A distinct advantage of the strong field approximation is that
it can in principle be used for an elliptically polarized driving field, whereas our 1D model is restricted to linear polarization.
On the other hand, ours is an exact solution of a well-defined system, and as such it is valid throughout the whole
frequency spectrum. Unlike the strong field approximation, it accounts for ``all electron trajectories'' not only those that give
rise to the harmonic radiation. In particular, the low-frequency components of the nonlinear current response affect the
propagation of the driver pulse through ionization, de-focusing by freed electrons, ionization losses,
and the nonlinear focusing.

\medskip
{\em The nonlinear Kerr effect}
\medskip

The $\delta$-potential atomic model only incorporates the nonlinear contribution of ground state to continuum transitions.
To capture the nonlinear contribution of virtual ground state to bound state electronic transitions we include a term
representing the nonlinear Kerr effect of the form
\begin{equation}
P(t) = 2 \epsilon_0 n_b \bar n_2 E^2(t)E(t)  ,
\end{equation}
where $\bar n_2$ is the nonlinear index and $n_b$ is the linear refractive
index. This effect enters the propagation equation via the nonlinear polarization term,
and gives rise to self-focusing and a cascade of lower-harmonic radiation.

\section{Illustrations involving near-threshold high-harmonic generation}

\smallskip
{\em HHG in a Xenon gas jet}
\smallskip

\begin{figure}[ht]
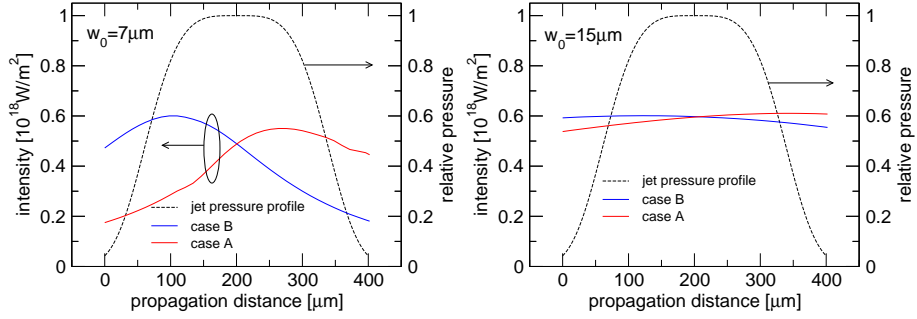

\centerline{
\includegraphics[clip,width=6cm]{figure3a.eps}
\includegraphics[clip,width=6cm]{figure3b.eps}
}
\caption{
For the sake of comparison, we examine four cases of focusing
geometry. Here, A and B denote geometric focus after and before the
center of the gas jet. The nominal maximal intensity
is kept constant.
}
\label{fig:3}
\end{figure}

The first illustrative example involves near-threshold HHG in a Xenon gas jet.  The incident pulse in each case is Gaussian in space
and time with pulse duration $t_p=85$ fs and center wavelength of $800$ nm, and focused spot size $w_0$.  We model the Xenon gas jet
as a vertical column which is tapered along the propagation or z-axis according to the jet pressure profile indicated by the dotted
line in Fig.~\ref{fig:3}.  This pressure or density profile has a constant region of length $L=200~\mu$m between $z=100-300~\mu$m and
tapers off on each side of this region.  For the examples in this paper we used 20~Torr for the maximal pressure in the gas jet.
Furthermore we consider the two distinct focusing cases where the input beam is focused
at $z=100~\mu$m close to the entrance  to the gas jet, and also the case that the input beam is focused at $z=300~\mu$m close to the
exit of the gas jet (We use this descriptive nomenclature of exit and entrance to the gas jet with the caveat that HHG can arise
before and after the exit and entrance due to the tapers in the gas density).  The variation of the on-axis intensity along the
z-axis is illustrated in Fig.~\ref{fig:3} for the cases of a) a narrow beam with $w_0=7~\mu$m, and b) a wide beam with $w_0=15~\mu$m.

In what follows we shall present examples of fully resolved simulations of pulse propagation and harmonic generation.
As a means to present our results we have chosen to plot the angularly resolved spectra for selected harmonics as
they propagate through the gas jet.  Such spectra have previously been utilized in both numerical simulations and were
measured experimentally~\cite{Schapper:10,Teichman2012}.  
For example, with reference to Figs.~4,6,7 each panel shows a color encoded plot of the spectral
intensity (red being maximum and blue minimum) as a function of the transverse wavenumber representing angular spread along
the horizontal axis, the center corresponding to zero propagation angle, and frequency along the vertical axis for the
harmonic indicated.  In particular, the frequency axis in each panel is centered on the harmonic order and has a spread
of one half of the harmonic spacing.  Each set of results is arranged into a $(3\times 3)$ array of panels, the panels are
arranged vertically according to the propagation distance into the gas jet, the top panel being the entrance and the bottom
panel the exit of the gas jet, and the panels are arranged horizontally from left to right for harmonic orders 9,11 and 13.

\bigskip

{\em Tighter focus case}

\medskip

\begin{figure}[ht]
\centerline{
a) Focus at gas jet exit
\hfill
b) Focus at gas jet entrance
}
\centerline{
\includegraphics[clip,width=6.5cm,height=6.5cm]{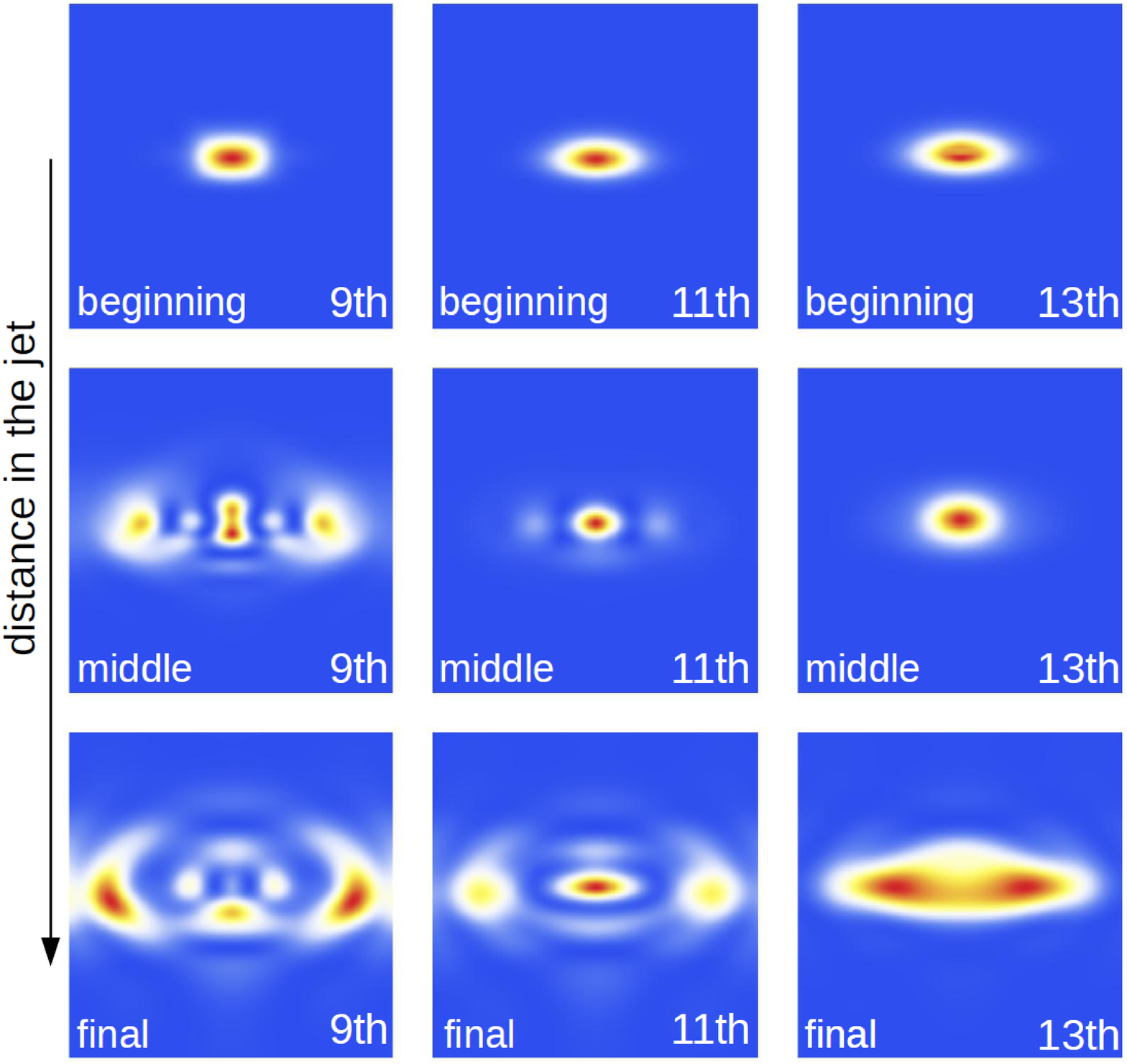}
%\hspace{-1cm}
\includegraphics[clip,width=6.5cm,height=6.5cm]{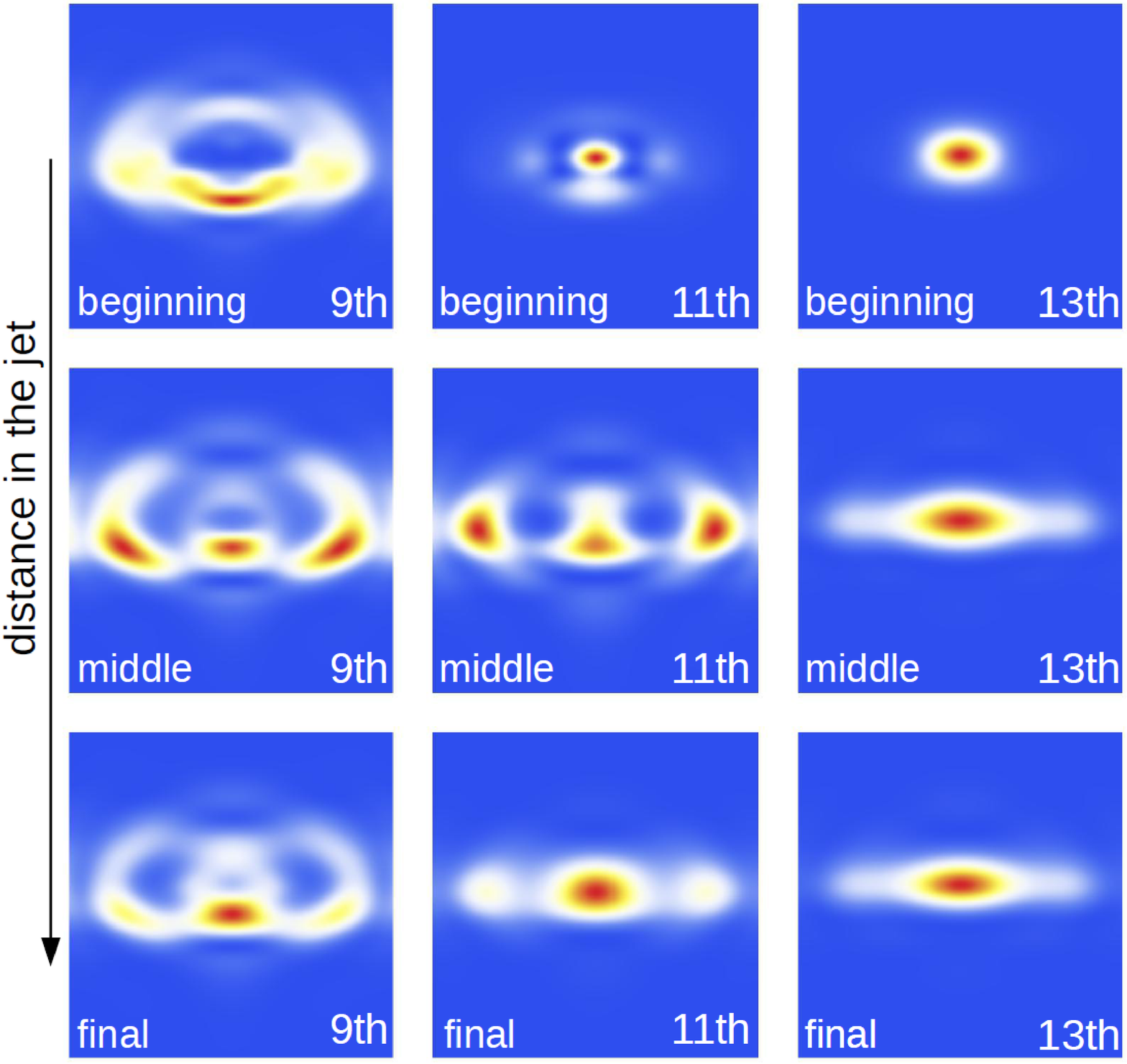}
}
\caption{
Angularly resolved spectra of the 9th, 11th, and 13th harmonics at three
different propagation distances through the gas jet for $w_0=7~\mu$m. Each panel shows a
frequency region (vertical axis) corresponding to one half of the harmonic
order, centered at the given harmonic frequency. The horizontal extent of each
panel corresponds to the transverse wave-number and thus represents the angle
of propagation.
}
\label{fig:4}
\end{figure}
\begin{figure}[ht]
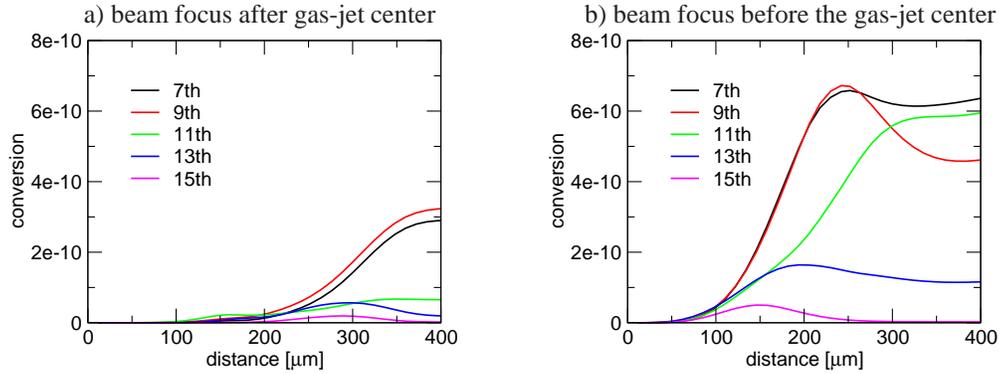

\centerline{
\hspace{0.9cm}
a) beam focus after gas-jet center
\hfill
b) beam focus before the gas-jet center
}

\centerline{
\includegraphics[clip,width=6cm]{figure5a.eps}
\hspace{1cm}
\includegraphics[clip,width=6cm]{figure5b.eps}
}
\caption{
Evolution of the conversion efficiency versus propagation distance in
the gas jet for two different beam focusing conditions. The gas density
is a smooth flat-top profile with a roughly constant pressure region
between 100 and 300 micron. The pulse intensities are comparable in both
cases, with their maxima approximately located at $z=300$ and $z=100$ micron
in case a) and b), respectively.
}
\label{fig:5}
\end{figure}

For the first example we chose a focused spot size of $w_0=7~\mu$m giving a Rayleigh range of $z_0=\pi w_0^2/\lambda=192~\mu$m for
the input field. In this case $L\simeq z_0$ and we expect that the focusing of the beam will impact phase-matching of the HHG via
the Gouy phase-shift incurred by the Gaussian beam.  The calculated angularly resolved spectra are shown in Fig.~\ref{fig:4}
for (a) focusing at the exit, and (b) focusing at the entrance of the gas jet, and we note that there is a marked
difference between the angularly resolved spectra for the two cases.  This is consistent with general expectations based on
phase-matching~\cite{lhuillier1991,salieres1995,constant_optimizing_1999}.  
For example, on-axis phase-matching requires cancellation between the accumulated Gouy phase-shift and the accumulated
atomic phase-shift due to the quantum path taken by the electron.  The Gouy phase-shift is a positive quantity whereas the quantum
phase is proportional to the z-derivative of the incident beam intensity due to the dependence of the quantum phase on the electron
quiver energy, which is positive for case (a) with the beam focused at the exit, and negative for the case (b) with the beam focused
at the entrance to the gas jet.  Thus on-axis phase matching is not a possibility for the case (a) with the input beam focused at
the exit, whereas phase-matching is a possibility in case (b) with focusing at the entrance to the gas jet.  The difference in the
angularly resolved spectra evident for cases (a) and (b) in Fig.~\ref{fig:4} may thus be traced to expected differences in
phase-matching conditions~\cite{salieres1995}.

If we further examine the angularly resolved spectra at the exit of the gas jet, shown as the lower set of panels in Fig.~\ref{fig:4}
for cases (a) and (b) we observe further consequences of the different phase-matching conditions: For case (a) for which strict
on-axis phase-matching is not possible we see that the off-axis HHG emission is more significant with respect to the on-axis
 emissions for each harmonic order
than for case (b) where on-axis phase-matching is possible. This is particularly pronounced for the $13^{th}$ harmonic where
case (a) is dominated by off-axis emission whereas case (b) is dominantly on-axis emission.  So the results of our simulations
are compatible with expectations based on on-axis phase-matching~\cite{lhuillier1991,salieres1995}.

\begin{figure}[ht]
\centerline{
\includegraphics[clip,width=8cm]{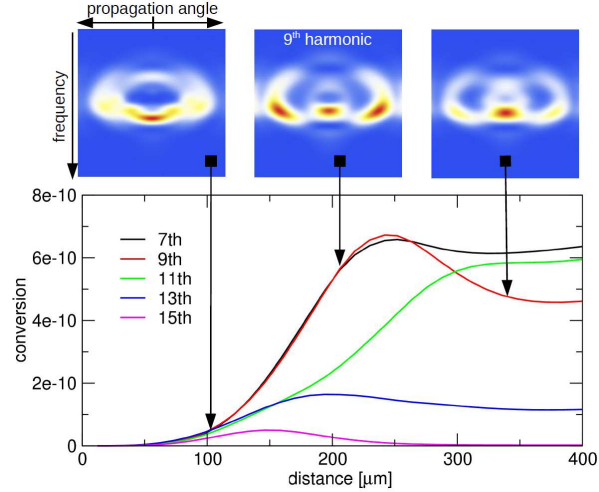}
}
\caption{
On-axis and off-axis propagating harmonic components exhibit different
phase-matching behavior. The top panel show the angularly resolved spectra
of the 9th harmonic at the beginning, in the middle and at the end of
the gas jet corresponding to case (b) from Fig.~\ref{fig:4}. Energy accumulates preferentially in the off-axis (conical)
components during the propagation in the center of the jet. Later, the
on-axis component becomes relatively stronger, as the pulse starts to
exit from the jet and the gas pressure it experiences starts to decrease.
This behavior correlates with the evolution of the energy accumulated in the
9th harmonic (shown as the red line in the lower panel).
Note that this is the harmonic which suffers the strongest losses.
}
\label{fig:6}
\end{figure}

To continue our discussion Fig.~\ref{fig:5} shows the evolution of the conversion efficiency for a variety of marked harmonics
during the propagation through the gas jet. (The conversion efficiency is obtained by integrating the angularly resolved spectrum
over the transverse wavenumber plane for a given harmonic order.)  It compares the cases of
(a) focus at the exit (left), and focus at the entrance to the gas jet (right), and shows that
the efficiency is higher in the latter case, as is generally accepted to be the case~\cite{salieres1995,Gaarde2008}. Furthermore, Fig.~\ref{fig:4} reveals that
the evolution of the harmonics during the pulse propagation through the jet is distinctly non-monotonic, that is, different
angular portions of the harmonic fields dominate at different propagation distances.  To illustrate that our simulation capability
can capture the complex interplay between HHG processes and phase-matching considerations
due to variation in the incident field and gas jet pressure, Fig.~\ref{fig:6} plots
the evolution of the harmonic conversion efficiency for several harmonics, the upper panels showing the evolution of the angularly
resolved spectrum for the $9^{th}$ harmonic at three propagation distances.  At the shortest and longest propagation distances note
that the spectrum has its maximum on-axis, whereas at the on-axis and off-axis become comparable in the middle.  It is interesting
that the turnaround in the conversion efficiency for the 9th harmonic that occurs for a propagation distance of around $200~\mu$m is
therefore dominated by a reduction in the off-axis as opposed to the on-axis emission.  This highlights the point that the spatial
evolution of the various harmonics involves a complex interplay between on-axis and off-axis emissions, and it is not generally
sufficient to assess the conversion efficiency based on the on-axis phase-matching considerations.

\bigskip

{\em Weaker focus case}

\medskip

\begin{figure}[ht]
\centerline{
a) beam focus after gas-jet center
\hfill
b) beam focus before the gas-jet center
}
\centerline{
\includegraphics[clip,width=6.5cm,height=6.5cm]{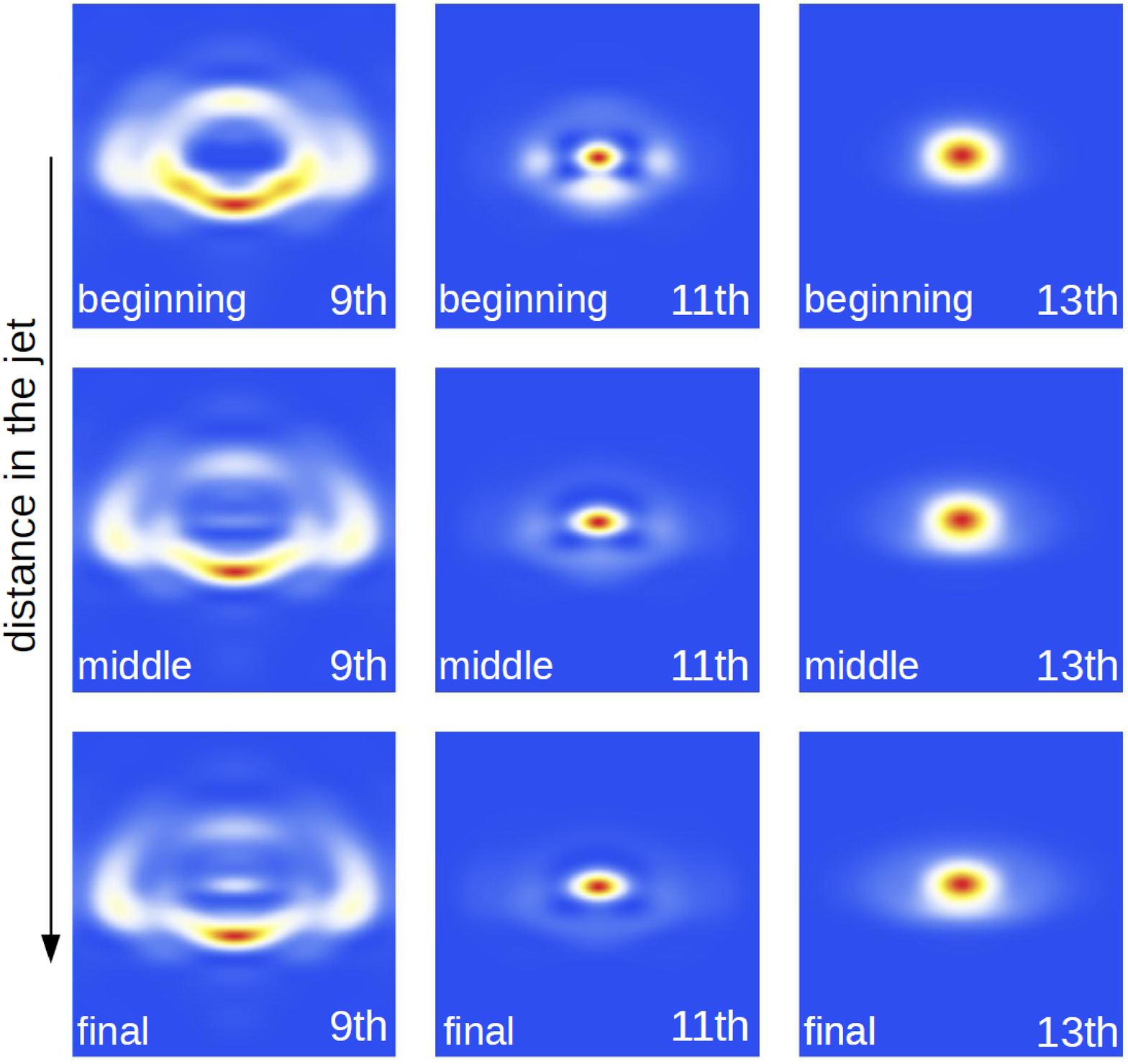}
%\hspace{-1cm}
%\includegraphics[clip,width=8cm]{figure7b.pdf}
\includegraphics[clip,width=6.5cm,height=6.5cm]{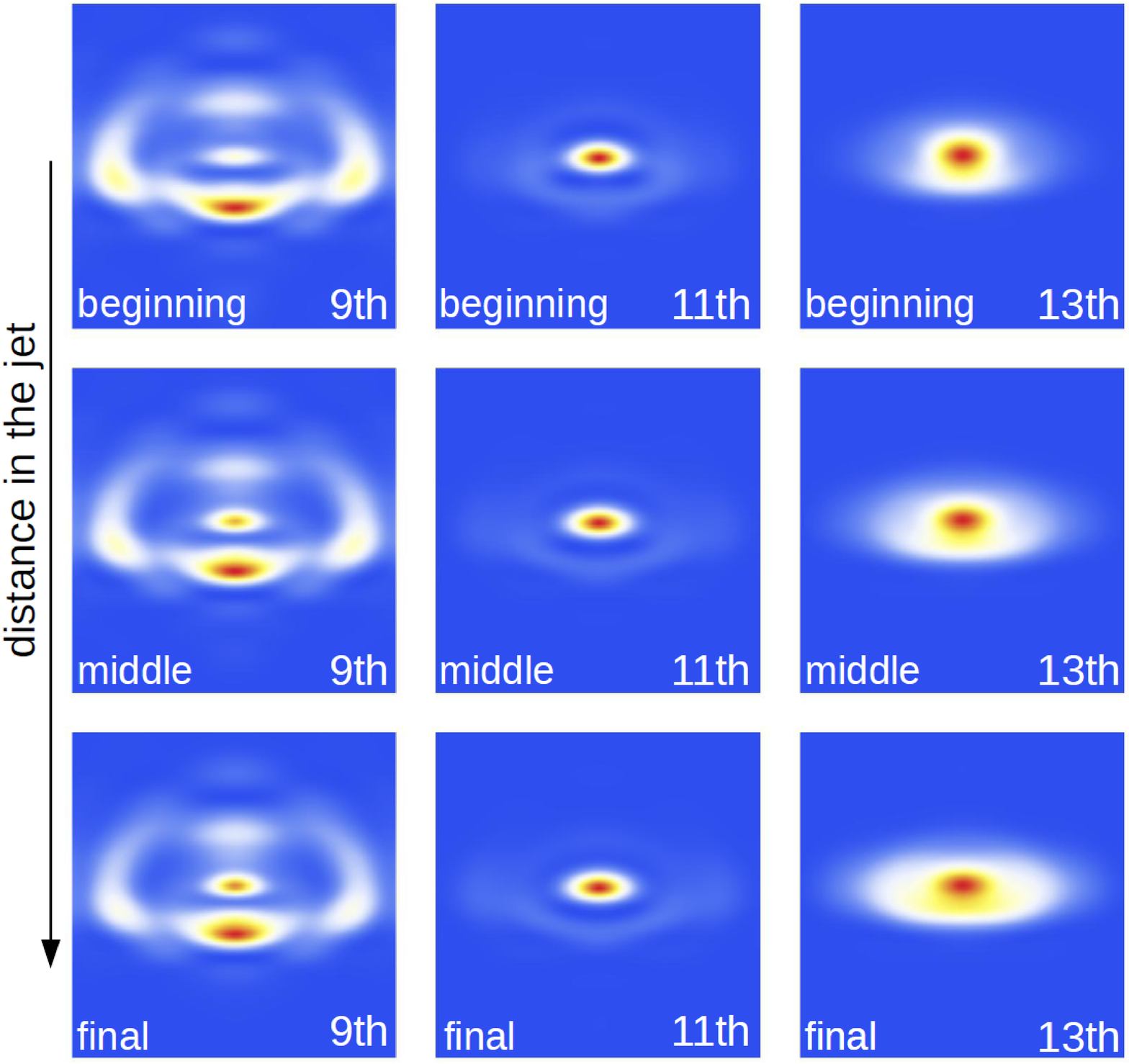}
}
\caption{
As expected, for a wider beam ($w_0=15\mu$m), the precise location of the beam focus is less important.
Angularly resolved spectra are shown for the 9th, 11th, and 13th harmonics at three different
propagation distances through the gas jet. Each panel shows a frequency region (vertical
axis) corresponding to one half of the harmonic order, centered at the given harmonic frequency.
The horizontal extent of each panel corresponds to the transverse wave-number and
thus represents the angle of propagation.
}
\label{fig:7}
\end{figure}

For this case we chose a focused spot size of $w_0=15~\mu$m giving a Rayleigh range of $z_0=\pi w_0^2/\lambda=883~\mu$m for the input
field. Now, $L<<z_0$ and we expect that the focusing of the beam will not impact phase-matching of the HHG in a major way.
The calculated angularly resolved spectra for this case are shown in Fig.~\ref{fig:7} for the cases of (a) focusing at the exit, and (b) focusing at the entrance of the gas jet, and we note that there is no major differences difference between the angularly resolved spectra for the two cases.  There are
small quantitative differences though, in particular in the visibility of the ring structure (cf. the 11th harmonic far-field patterns).

\begin{figure}[ht]
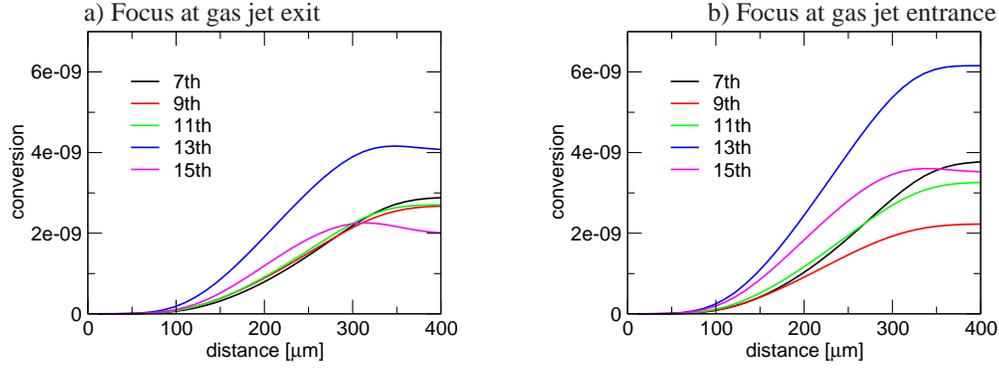

\centerline{
\hspace{0.9cm}
a) Focus at gas jet exit
\hfill
b) Focus at gas jet entrance
}

\centerline{
\includegraphics[clip,width=6cm]{figure8a.eps}
\hspace{1cm}
\includegraphics[clip,width=6cm]{figure8b.eps}
}
\caption{
Evolution of the conversion efficiency versus propagation distance in
the gas jet for two different beam focusing conditions. The gas density
is a smooth flat-top profile with a roughly constant pressure region
between 100 and 300 micron. The pulse intensities are comparable in both
cases, with their maxima approximately located at $z=300$ and $z=100$ micron
in case a) and b), respectively.
}
\label{fig:8}
\end{figure}

Figure \ref{fig:8} provides a comparison of the conversion efficiency evolution versus the
propagation distance for the two focusing cases.  This figure shows that while the optimal position of the beam
focus is still at the entrance into the gas jet, as generally accepted~\cite{salieres1995}, 
for weaker focusing of the input field the qualitative behavior
is relatively insensitive to the position of the input beam focus with respect to the gas jet.

Finally we would like to point to one other feature from our simulations that correlates with previous studies: Inspection of the
high resolution image for the angularly resolved spectra for the 11th harmonic in Fig.~\ref{fig:7} reveals that the output
spectra in the bottom row display ring structures.  Such structures have previously been seen experimentally and appeared
in propagation simulation in (3+1) dimensions employing the strong-field approximation~\cite{salieres1995,Schapper:10}.  
Moreover the appearance of these angle-frequency spectral features has been associated with the interference between 
the short and long path contributions to HHG.  
%XXX modifications
Recently, Teichman et al.\cite{Teichman2012} demonstrated,
experimentally and numerically, that rings  in the angle-frequency spectra
can be attributed to the interferences in the single-atom response in the transverse cross-section
profile, and that their appearance depends on particular conditions such as intensity
and specific harmonics.
%XXX modifications
It is satisfying that our computational approach can
reveal such structures without resort to the strong field approximation.
%XXX modifications
Moreover, inspection of near-field cross-sections (not shown) 
of the frequency-filtered atomic response shows that complex far-field patterns
correlate with the complex spatial transverse distribution  of the HHG ``source.'' 
This suggests, in keeping with~\cite{Teichman2012}, that dynamics which controls the transverse
properties of the atomic response play equally important role in macroscopic 
selection through phase matching.
%XXX modifications

\smallskip
{\em HHG in a femtosecond enhancement cavity}
\smallskip

As a final illustration, we show results for harmonic generation in a
xenon gas jet placed inside a passive femtosecond enhancement cavity
(fsEC).  This experimental configuration has been demonstrated as a
means to generate XUV frequency combs with the harmonic light being
generated at each cavity round trip \cite{Jones05, Gohle05}.  In this
geometry the ionized gas jet behaves as a nonlinear medium at the
focus of a 6 m ring cavity.  A pair of 15 cm radius of curvature
focusing mirrors lead to an intracavity spot size of $w_0 = 15\ \mu$m
within the jet.  The fsEC has a 1\% input coupler that allows pulse
energy enhancement over 200 times relative to the incident pulse
train leading to peak intensities in excess of $1 \times 10^{14}$
W/cm$^2$ when the gas jet is off.  The harmonics are coupled out of
the cavity using a thin sapphire plate, aligned at Brewster's angle for the
fundamental pulse, and resolved spatially by reflection from an XUV
grating.

For initial conditions in these simulations, we have utilized a
previously calculated temporal profile of the pulse just before it
enters the gas jet. This pulse profile was obtained from a 1D
steady-state calculation of the fundamental pulse building up inside
the fsEC in the presence of the xenon gas and includes the effects of
cavity dispersion.  The high nonlinearity of ionization leads to
chirped pulses circulating in the cavity and limits the achievable
intensities in this geometry.  Details of this model and its
implications for HHG in a fsEC can be found in Ref. \cite{Carlson11}.

Yet another modification of the current model consists in including the
surviving plasma in the jet based on the estimated levels calculated in \cite{Carlson11} . Because the 20 ns cavity round-trip time is too
short for the plasma to decay entirely, the pulse propagating through the jet
experiences defocusing due to electrons freed during the previous
passes.  These electrons have been liberated from their parent atom
for a sufficiently long time, and have equilibrated into a true plasma
(note that the free-electron states included in the quantum model that
drives the pulse propagator are of different nature: they did not have
enough time to thermalize, and their interactions is mainly with the
nearby parent ion).  Therefore, the influence of the plasma remnant
can be included within the linear chromatic properties of the
gas. Here we also assume that the diffusion resulted in a nearly
homogeneous spatial distribution of free electrons, and the
corresponding modification of the refractive index is constant over
the cross-section of the beam. Of course, the spectral nature of our
pulse solver allows to endow this susceptibility contribution with the
correct frequency dependence, $\chi(\omega) \sim
-\omega^2_{plas}/\omega^2$.

\begin{figure}[ht]
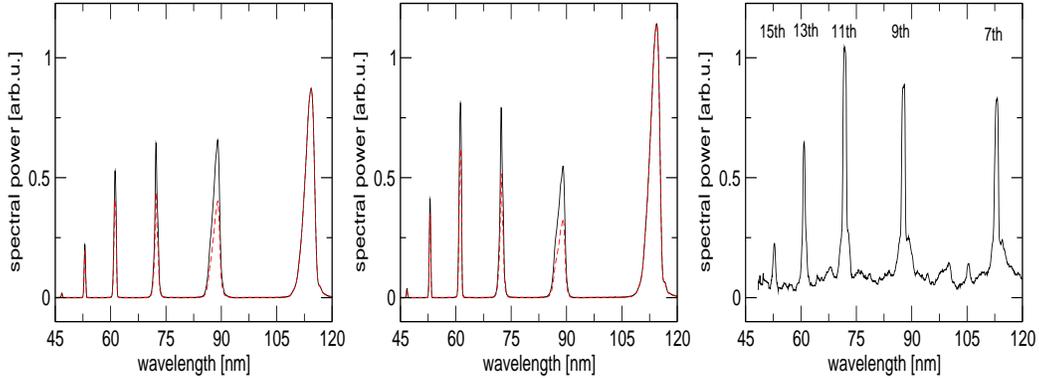

\centerline{
  \includegraphics[clip,width=4.5cm,height=5cm]{figure9a.eps}
  \includegraphics[clip,width=4.5cm,height=5cm]{figure9b.eps}
  \includegraphics[clip,width=4.5cm,height=5cm]{figure9c.eps}
}
\caption{
Harmonic spectrum calculated (left panels) for conditions reflecting those in the
femtosecond enhancement cavity. Dashed and full line compare results for
simulation with and without inclusion of losses. Left most panel: beam focus
at the ``entrance'' into the gas jet. Middle panel: beam focus at the
``exit'' from the gas jet. Experimental spectrum shown in the right
panel, with an account for the grating efficiency.
}
\label{fig:9}
\end{figure}

Figure~\ref{fig:9} shows simulated and experimental spectra of harmonics from
7th to 15th. While details of relative strength of harmonics 9th to 15th depend
on the exact placement of the beam focus with respect to the center of the jet
(compare the two panels on the left), and on the absorption included in our simulation
(compare full and dashed lines), we see that harmonics 11 and 13 are the most pronounced,
and that harmonic power starts to decrease at harmonics 15th (there are many more
harmonics generated, but are not discernible on the linear scale of this figure).
This is compatible with the spectrum recorded in the experiment. On the other hand,
the simulated 7th harmonic seems to be too strong. We think that this is mainly due to
the fact that the model susceptibility of the gas has no absorption at this wavelength
(this is due to limited frequency extent of the available data, see the red line in Fig.~\ref{fig:1}).
The effect of medium absorption is indeed
clearly visible in the 9th harmonic where our model absorption data exhibit a maximum.
These results make it thus evident that it is important to obtain
as realistic as possible a data set for both the index of refraction and
absorption of the gas.

We certainly do not expect that the simplistic one-dimensional quantum model
put forward here could reproduce experimental results in a one-to-one manner. However,
we see the above comparison as very encouraging --- it shows that the
model, especially when coupled with good-quality dispersion and absorption
data for the gas, can serve as a practical tool to study the qualitative trends that
govern the nonlinear interactions that span the frequency range from infrared to
high harmonics.

\section{Conclusion}

In conclusion, we have presented a new computational approach for femtosecond pulse propagation in the transparency
region of gases that permits full resolution in three space dimensions plus time while fully incorporating quantum
coherent effects such as high-harmonic generation and strong-field ionization in a holistic fashion.  The key innovation
that makes this possible is the use of 1D atomic model with delta-function potential to compute the nonlinear optical
response due to ground state to continuum transitions, and the closed form solution for the nonlinear response leads
to significant reductions in computing time compared to a direct solution of the atomic Schr\" odinger equation. In spite
of the approximation of a 1D atomic model we contend that our computational approach will be of great utility as a research
tool in a number of forefront areas.  For example, in the field of femtosecond filamentation in gases an outstanding
question is whether coherent light-matter interactions can play a significant role. The current models, based
upon individual ingredients of the Kerr effect, multi-photon-ionization, and a Drude model for the plasma, are coming
under increasing scrutiny.  Our new approach answers to this need and provides a computationally viable means of performing
simulations of filamentation which includes the quantum coherent atomic effects.  Moreover, it eliminates the part
of the filamentation model which is considered to be  the weakest link, namely ionization and concomitant effects
due to freed electrons. For this initial presentation we have chosen to illustrate our computational approach using
the example of HHG in a gas jet as this is a distinctly quantum coherent effect and also requites an enormous spectral
width.  In particular, using the example of near-threshold HHG in a Xenon gas jet we hope to have shown that our
computational approach yields results in keeping with general expectations, and that it is a useful tool to study
qualitative dynamic trends in a completely self-consistent way.  As a more concrete example we presented a qualitative
comparison between our model and results from an in-house experiment on extreme ultraviolet generation in a femtosecond
enhancement cavity.

\section{Appendix}

Here we summarize the formulas needed to evaluate the nonlinear component
of the current induced in the system described by the time-dependent Schr\"odinger
equation~(\ref{eq:TDSE}).

An arbitrary time-dependent external field $F(t)$ enters through the classical
electron trajectory $x_{cl}(t)$ (here we assume that $x$ is the direction of the
optical field polarization)
\begin{equation}
p_{cl}(t) =           - \int_0^t F(\tau )       d\tau \ \ \ \ , \ \ \ \
x_{cl}(t) = \phantom{-} \int_0^t p_{cl}(\tau )   d\tau \ \ .
\end{equation}
Assuming that the system was initially in its ground state,
an auxiliary quantity $A(t)$ is obtained first as a solution to the following
integral equation:
\begin{equation}
\label{eq:IEqnA}
A(t)  =  \psi_{R}(-x_{cl}(t), t) +
\frac{i B}{\sqrt{2 \pi i }} \int_0^t{\! dt' \frac{e^{+i \frac{B^2}{2} (t'-t) }}{\sqrt{t-t'}}
       \exp{\left[ \frac{i (x_{cl}(t)-x_{cl}(t'))^2}{2 (t-t')} \right]} A(t')}\ ,
\end{equation}
where the right-hand-side is
\begin{equation}
\label{eq:rhs}
\psi_{R}(x,t) \equiv
    \frac{e^{+B x}}{2} {\rm erfc}\! \left(\! \frac{i B t + x}{\sqrt{2 i t}}\! \right)
 +  \frac{e^{-B x}}{2} {\rm erfc}\! \left(\! \frac{i B t - x}{\sqrt{2 i t}}\! \right).
\end{equation}
Having calculated $A(t)$, the nonlinear (in the strength of the external field $F$) current
is expressed as a sum
\begin{equation}
J^{(nl)} = J_{SS}^{(nl)} + J_{FS}^{(nl)} \ \ ,
\end{equation}
with the components listed below:
\begin{equation}
J_{SS}^{(nl)}
= 2 {\rm Im}\! \left\{\! \int_0^t \! \! \! \!dt_1 \! \! \int_0^{t_1} \! \! \! \! \! \!dt_2
\frac{(-i)^{\frac{3}{2}} B^3 }{\sqrt{2\pi}} \frac{e^{+i \frac{B^2}{2} (t_2-t_1) }}{\sqrt{t_1-t_2}}
\left[ e^{\frac{i [x_{cl}(t_1)- x_{cl}(t_2)]^2}{2(t_1-t_2)}}
A^*(t_1) A(t_2) - 1 \right]
\frac{ x_{cl}(t_1) - x_{cl}(t_2)}{ t_1-t_2}
\! \right\}
\label{eq:JSSnl}
\end{equation}
\begin{eqnarray}
J_{FS}^{(nl)}
&=&{\rm Im}\left\{ i B^3\!\!\! \int_0^t\!\!\! dt_1 A^*(t_1)
  e^{+B x_{cl}(t_1)} {\rm erfc}\!\left(\frac{(1\! +\! i)(B t_1 - i x_{cl}(t_1))}{2 \sqrt{t_1}}\right)
\right\} \cr
&-&{\rm Im}\left\{ i B^3\!\!\! \int_0^t\!\!\! dt_1 A^*(t_1)
  e^{-B x_{cl}(t_1)} {\rm erfc}\!\left(\frac{(1\! +\! i)(B t_1 + i x_{cl}(t_1))}{2 \sqrt{t_1}}\right)
\right\} \cr
&-&{\rm Im}\left\{2 B^3\!\!\! \int_0^t\!\!\! dt_1 x_{cl}(t_1)
\left(
i B\ {\rm erfc}\! \left(\frac{(1\! +\! i)B \sqrt{t_1}}{2}\right)
- \frac{1+i}{\sqrt{\pi t_1}} e^{-i\frac{B^2}{2} t_1}
\right)
 \right\}
\label{eq:JFSnl}
\end{eqnarray}

Details of the derivation and description of numerical implementation can be found
in Ref.~\cite{Brown2011}. An open-source implementation is also available
upon request from the authors (M.K. or J.B.).

\medskip

\noindent{\bf Acknowledgments: }
M.K. and E.M.W. acknowledge support from the
U.S. Air Force Office for Scientific Research, through the MURI grant FA9550-10-1-0561.
The parallel pulse propagation framework used in this study was developed by J.A. with the funding
from AFOSR grant FA9550-11-1-0144. R.J.J, E.M.W, and D.R.C. are supported from AFOSR grant FA9550-12-1-0048.

\end{document}